\documentclass[pr,onecolumn,superscriptaddress,showpacs]{revtex4}
\usepackage{graphicx,bm,amsmath,mathrsfs,amsfonts,color,amssymb,latexsym,revsymb,amsthm,color,enumerate,hyperref,algorithm}
\usepackage{algpseudocode}
    \renewcommand{\algorithmicrequire}{\textbf{Input:}}
    \renewcommand{\algorithmicensure}{\textbf{Output:}}
\usepackage{multirow}

\newcommand{\bS}{{\bm S}}
\newcommand{\Tgate}{{\rm T}}

\begin{document}

\title{Quantum-circuit design for efficient simulations of many-body quantum dynamics}

\author{Sadegh Raeisi}
\affiliation{Institute for Quantum Information Science, University of Calgary, Alberta T2N 1N4, Canada}
\affiliation{Institute for Quantum Computing, University of Waterloo, Ontario N2L 3G1, Canada}
\author{Nathan Wiebe}
\affiliation{Institute for Quantum Information Science, University of Calgary, Alberta T2N 1N4, Canada}
\affiliation{Institute for Quantum Computing, University of Waterloo, Ontario N2L 3G1, Canada}
\author{Barry C. Sanders}
\affiliation{Institute for Quantum Information Science, University of Calgary, Alberta T2N 1N4, Canada}

\begin{abstract}
We construct an efficient autonomous quantum-circuit design algorithm for creating efficient quantum circuits
to simulate Hamiltonian many-body quantum dynamics for arbitrary input states.
The resultant quantum circuits have optimal space complexity and
employ a sequence of gates that is close to optimal with respect to time complexity.  We also devise an algorithm that exploits commutativity to optimize the circuits for parallel execution.
As examples, we show how our autonomous algorithm constructs circuits for simulating the dynamics of
Kitaev's honeycomb model and  the Bardeen-Cooper-Schrieffer model of superconductivity. Furthermore we provide numerical evidence that the rigorously proven upper bounds for the simulation error here and in previous work may sometimes overestimate the error by orders of magnitude compared to the best achievable performance for some physics-inspired simulations.
\end{abstract}

\maketitle

\section{Introduction}
Feynman proposed quantum computing as a means to ``imitate'' quantum dynamics
in order to overcome apparent intractability of universal quantum simulation on classical computers~\cite{Fey82}.
He conjectured that a universal quantum simulator (UQS) could efficiently simulate quantum evolution.
Lloyd formalized Feynman's concept by employing a Trotter ordered-operator expansion to convert continuous time evolution 
into a quantum circuit~$\mathcal C$ comprising unitary quantum gates~\cite{Llo96}.
The UQS is now also referred to as ``digital quantum simulation'',
both theoretically~\cite{ADLH+05,WML+10,WMBL11,WBA11} and experimentally~\cite{LHN+11}.
The adjective ``digital'' is used to contrast with the term ``analogue quantum simulation'',
which aims to emulate evolution of a Hamiltonian~$\hat{H}$ in a custom--designed
experiment~\cite{DDL03,ABVC09,GKZ+10,BN09,FSGPS08}.
The importance and near-future feasibility of the (digital) UQS,
albeit without quantum error correction,
drives experimental efforts to create these simulators for restricted types of $\hat{H}$~\cite{LHN+11}.

We present the first autonomous algorithm to design circuits for simulating the evolution generated
by a general $n$-qubit $k$-local Hamiltonian $\hat{H}^{(n)}$
within a pre-specified tolerance~$\epsilon$.
An $n$-qubit $k$-local $\hat{H}^{(n)}$ is defined to be a linear combination of~$m$ Hamiltonians~$\hat{\mathfrak{h}}_j^{(n)}$,
each acting on $n$ qubits
as an identity operator~$\openone$ on all but $k\in {\rm polylog}(n)$ qubits~\cite{KSV02},
and polylog$(n)$ is a polynomial function of $\log n$.
Our tolerance~$\epsilon$ is the worst-case $2$--norm distance between the true evolved state
under the specified evolution and the simulated output state, maximized over all allowed input states.  We also show in Sec.~\ref{appendix:numest} that these worst--case error bounds that go into these estimates can overestimate the error for random $2$--local Hamiltonians by orders of magnitude, which suggests that UQS experiments may be much more feasible than previous simulation work has suggested~\cite{WBHS10,WBHS11,PZ12,BACS07a,BACS07b}.

In our analysis, we consider two independent cases of universal gate sets.
One case corresponds to a finite set comprising a single two--qubit entangling gate plus a finite number of one--qubit
gates.
The second case incorporates a single two--qubit entangling gate plus both discrete and continuously-parameterized single--qubit gates.
Strictly speaking only a finite gate set should be permitted for quantum error correction and scalability,
but there is a trend in experimental studies to report quantum simulation with continuously-parametrized single-qubit gates.
We want our algorithm to be relevant both to the strict case of a finite gate set and to experimental efforts that
employ continuous tunability.

Our resultant circuits are not
only efficient (meaning that the circuit size scales polynomially with the number of simulated qubits for fixed $k$) but also uses the smallest number of qubits possible given the size of the system being simulated.  
Additionally, the circuit size also scales near--optimally with the run-time $t$ of the simulation.

This minimization over space and time costs
(number of qubits required in the simulator and number of gates)
is important for making quantum simulators as close as possible to practical implementation.
Each additional qubit and each additional gate can be challenging to implement in practice so reducing
these costs is not only important for proving that the scaling is efficient hence possible in principle
but also to reduce the costs to make the simulation feasible with small simulators in the near future.
Our minimum run-time algorithm is also improved by parallelizing gates by grouping commuting terms in the 
Trotter decomposition of the evolutionary operator, thus enhancing the near-term feasibility of the quantum simulator.
Our work thus enables feasibility of UQS circuits.
Experimental UQS circuits will be valuable to predict resultant states under $\hat{H}$-evolution
or to provide UQS-generated states
as inputs to quantum algorithms for purposes such as acquiring spectral properties or eigenstates of Hamiltonians~\cite{AL97}
or to determine particle scattering, e.g.\ in a relativistic quantum field theory~\cite{JLP11a,JLP11b}.
The ground state is regarded as the most important eigenstate as it uniquely determines all properties of the system~\cite{HK64}
and  could be used to solve outstanding condensed-matter problems such as determining the energy gap
for general Bardeen-Cooper-Schrieffer (BCS) superconductivity models of \emph{finite} systems~\cite{WBL02}.
Dynamical simulation algorithms have been devised to simulate ground-state properties
via adiabatic state preparation~\cite{WBL02} or via dissipative interaction with the environment~\cite{WML+10,WMBL11}.

\section{Algorithm for designing the quantum simulator circuit}
\label{sec:algorithm}

\subsection{Concept}
\label{subsec:concept}
Our algorithm yields a string representing a quantum circuit that comprises a sequence of quantum gates to simulate $k$--local Hamiltonian
evolution within fixed 2-norm distance $\epsilon$.
The  $n$-qubit $k$-local Hamiltonian is expressed in the Pauli operator basis as
\begin{equation}
\label{eq:Hsum}
	\hat{H}^{(n)}=\sum_{j=1}^m a_j\hat{\mathfrak{h}}_j^{(n)},\;
	 \hat{\mathfrak{h}}_j^{(n)}=\otimes_{\ell=1}^{n}\hat{\Xi}^{(n)}_{j\ell},
\end{equation}
with $\Xi^{(n)}_{j\ell}\in\{\openone,X,Y,Z\}$ such that
\begin{equation}
	X=\left[\!\begin{array}{cc}0&1\\1&0\end{array}\!\right],
	Y=i\left[\!\begin{array}{cc}0&\!\!-1\\1&0\end{array}\!\right],
	Z=\left[\!\begin{array}{cc}1&0\\0&\!\!-1\end{array}\!\right].
\end{equation}
The total number of non-identity Pauli operators in each $\hat{\mathfrak{h}}_j^{(n)}$
is at most an $n$-independent constant $k$.

Our algorithm is designed to produce a poly($n$)--size description of a quantum circuit that implements a unitary $\tilde{U}^{(n)}(t)$ such that
\begin{equation}
	\left\|\exp\left(-i \hat{H}^{(n)} t\right)- \tilde{U}^{(n)}(t)\right\| \le \epsilon,
\end{equation}
where $\|\cdot \|$ denotes the $2$-norm.
This condition implies that the trace distance between the ideal evolution and the simulated evolution is at most $\epsilon$ for any initial state~\cite{BACS07a,BACS07b}.  
Below we discuss the algorithmic input, processing and output.

\subsection{Input}

The algorithm requires the following inputs:
\begin{itemize}
\item[] $n$: number of qubits in the system;
\item[] $k$: locality parameter of the Hamiltonian;
\item[] $t$: evolution time for the simulation;
\item[] $\epsilon$: worst-case 2-norm error tolerance (distance) between the true evolved state and the simulated state;
\item[] $\varpi$; specifies which single-qubit gate set to use, namely $\{H,\Tgate=Z^{-1/4}\}$
	or the continuously parametrized set $\{H,R_z(\theta)\}$, with $Z=R_z(\pi/2)$;
\item[] $\left[\hat H^{(n)}\right]$: bit-string representation of the $n$-qubit $k$-local Hamiltonian.
\end{itemize}
The Hamiltonian~$\left[\hat H^{(n)}\right]$ is entered into the algorithm as a bit string no larger than
a poly($n$)--size representation for this input to be efficient.
Our algorithm accepts the Hamiltonian~$\hat{H}^{(n)}$ input as the bit-string representation
\begin{equation}
\label{eq:Hstring}
	\left[\hat H^{(n)}\right]:=\left\{\left(a_j,\bm{l}_j,\bS_j\right);j=1,\ldots,m\right\},
\end{equation}
with
\begin{equation}
\label{eq:l_j}
	\bm{l}_j=\left(l_{Xj},l_{Yj},l_{Zj}\right)
\end{equation}
the vector corresponding to the numbers of
each type of Pauli operators  in $\hat{\mathfrak{h}}_j^{(n)}$
and
\begin{equation}
\label{eq:S_j}
	\bS_j=\left(\bS_{Xj},\bS_{Yj},\bS_{Zj}\right)
\end{equation}
with $\bS_{Xj}$, $\bS_{Yj}$ and $\bS_{Zj}$
the strings corresponding to the positions of each of the $X$, $Y$ and $Z$ operators respectively in~$\hat{\mathfrak{h}}_j^{(n)}$.

In Eq.~(\ref{eq:Hstring}), substrings representing Hamiltonians $\hat{\mathfrak{h}}^{(n)}_j$ appear as triplets of strings.
These triplets are delimited by parentheses thus ensuring easy parsing of the overall string for $\left[\hat H^{(n)}\right]$.
Whereas a general matrix representation of~$\hat{H}^{(n)}$ is exponentially large in~$n$,
our representation~$[\hat H^{(n)}]$ of this $k$-local Hamiltonian has poly($n$) size.

As an example of this encoding, consider the three--qubit Hamiltonian
\begin{equation}
\label{eq:Hn}
	\hat H^{(n=3)}
		=\: X\otimes X\otimes \openone +2\: Y\otimes Y\otimes \openone +4\: Y\otimes \openone \otimes Z
\end{equation}
with $m=3$.
This Hamiltonian is representated as
\begin{align}
	\bm{a} =& (1,2,4),\nonumber\\
	\bm{l}_1=& (2,0,0),\;\bS_{X1}=(1,2),\nonumber\\
	\bm{l}_2=&(0,2,0),\;\bS_{Y2}=(1,2),\nonumber\\
	\bm{l}_3=&(0,1,1),\;\bS_{Y3}=(1), \bS_{Z3}=(3),
\end{align} 
and all other strings are empty.

Our algorithm can compute the number of time steps~$r$ of equal duration~$\frac{t}{r}$ 
so that the evolution from~$0$ to~$t$ is broken down into a sequence of finite steps.
Furthermore our algorithm employs the  Trotter--Suzuki (TS) ordered-exponential decomposition~\cite{Suz90,Suz91,BACS07a,BACS07b} of order~$\chi$
to simulate the evolution sequentially over each of the~$r$ time steps.
However, our algorithm determines~$r$ and~$\chi$ based on optimizing an upper bound on tolerance,
and superior choices of~$r$ and~$\chi$ are possible but not by using our algorithm or any other known algorithm.  We provide numerical evidence of this fact, for random $2$-body Hamiltonians, in Section~\ref{appendix:numest}.

As the user of our algorithm might wish to employ or test other choices of~$r$ and~$\chi$,
we design the algorithm so that the user can override our choices of~$r$ and~$\chi$.
If our algorithmically determined value of~$r$ is overridden,
the guarantee that the quantum simulations yields a resultant state within tolerance~$\epsilon$ no longer holds.
Therefore, overriding the algorithm's $r$ comes with the warning that the error in the quantum simulation is unknown.
Specifically setting $r=0$ and $\chi=0$ causes our algorithm to determine optimal $r$ and $\chi$ values based on minimizing
the number of gates required to guarantee that the simulated state has 2-norm error less than $\epsilon$ for any input state.
Positive values of~$r$ and~$\chi$ override our algorithmic determination of these parameters and use the input values instead.

The purpose of the input variable~$\varpi$ is to enable circuit design that strictly uses a finite gate set in 
accordance with principles of quantum error correction and scalability~\cite{NC00}
or to include a continuously-parametrized single-qubit gate, namely rotations by~$\theta$ around the $z$-axis
using unitary operator~$R_z(\theta)$, in accordance with prevalent experimental practice~\cite{LHN+11}.
More generally one could consider the case that~$\varpi$ is any desired universal gate set,
but here we restrict our attention to just two single-qubit gates,
either~$\{H,T=Z^{-1/4}\}$ or $\{H,R_z(\theta)\}$,
so~$\varpi$ is a binary, or logical, variable.
In our algorithm, both of these sets are accompanied by two-qubit controlled-not gate
CNOT$_{ij}$ with the $i$ labelling the control qubit and $j$ labelling the target qubit.

\subsection{Output}
\label{subsec:output}

Our algorithm yields an output string~$[\mathcal{C}]$ that represents the resultant quantum circuit.
This quantum circuit is a sequence of quantum gates that simulate the dynamics of the quantum system
for an arbitrary input state.
In our algorithm the basic quantum gates are members either of the finite-size universal instruction set~$\{H,T,\text{CNOT}\}$
or the continuously-parametrized set ~$\{H,R_z(\theta),\text{CNOT}\}$.

The single qubit gates are represented in the output as a string of the form ``$Hx$'' or ``$\Tgate x$''
with~$x$ a bit string labeling the qubit acted upon by gate~$T$ or~$H$, respectively.
These gates, along with the identity~$\openone$, which is not explicitly stated in the circuit,
can be parallelized and concatenated to form circuits.
For example, the string $H1T2$ represents a circuit that first performs the Hadamard operation
on qubit $1$ then a $\pi/8$--gate on qubit $2$.
The CNOT operation is represented similarly with the string ``CNOT$x$,$y$'' representing the
quantum operation ${\rm CNOT}_{xy}$.
For the case of the gate set that includes continuously parameterized single--qubit z--rotation gates, $R_z(\theta)$, in the output,
The action of $R_z(\theta)$ on qubit $x$ is represented as the string $``RZ[\theta],x$'' where $[\theta]$ is a string representing  the rotation angle $\theta$.

\subsection{Processing}
\label{subsec:processing}
Our circuit-design algorithm proceeds through three stages.
\begin{enumerate}
\item[] Hamiltonian sorting algorithm:
	mutually commuting terms in~$H^{(n)}$ are grouped together resulting in parallelized quantum simulation
	that reduces total runtime.
\item[] Trotter-Suzuki algorithm:
	$U^{(n)}(t):=\exp\{-i\hat{H}^{(n)}t\}$ is decomposed into a sequence of exponentials of Pauli operators.
\item[] Circuit design algorithm for Pauli--exponentials:
	determine the quantum circuit and convert into output string~$[\mathcal{C}]$.
\end{enumerate}
These algorithmic stages are described in the following sections.

\subsection{Summary}
\label{subsec:summary}

An algorithm consists of input and output with the output obtained by processing the input. 
We have been careful to discuss the input and output as bit strings as our algorithm is classical and runs on a
classical computer. 
The output of the algorithm is a design procedure to construct a quantum simulator circuit that would simulate
the evolution of a quantum state.
In the following sections we describe the algorithmic stages.

\section{Trotter--Suzuki Formulas}\label{sec:TS}

In this section we discuss the second stage of the algorithm,
which concerns decomposing the TS ordered-operator exponential 
into a sequence of exponentials of tensor products
of Pauli operators.
The second stage of the algorithm is discussed first because the first stage groups these TS terms
so understanding the second stage helps to understand the terms being grouped in the first stage.

We use the TS method to determine a product of exponentials of $\hat{\mathfrak{h}}_j^{(n)}$ that
approximates $U^{(n)}(t)$ within 2-norm distance $\epsilon$~\cite{Suz90,BACS07a,BACS07b}.
The total time of evolution~$t$ is divided into~$r$ time intervals each of equal duration
\begin{equation}
\label{eq:Deltat}
	\Delta t = \frac{t}{r}.
\end{equation}
For $\tilde{U}^{(n)}_\chi$ the $\chi^\text{th}$-order TS iterate approximating the $n$-qubit unitary evolution~$U^{(n)}$,
the distance between the `true' evolution operator~$U^{(n)}(t)$ and the TS approximated evolution operator is
\begin{equation}
	\left\|U^{(n)}(t)-U^{(n)}_\chi\left(\Delta t\right)^r\right\|\in O\left(\frac{t^{2\chi+1}}{r^{2\chi}}\right).
\end{equation}

The iterative TS formula for generating $U_{\chi}$ is well known~\cite{Suz90}.
The formul\ae~are widely used in quantum simulation algorithms because they generate an approximation
\begin{equation}
	U^{(n)}_\chi\left(\Delta t\right)
		=	\exp\left\{-ia_{j_1}\hat{\mathfrak{h}}^{(n)}_{j_1}t_1\right\}
			\exp\left\{-ia_{j_2}\hat{\mathfrak{h}}^{(n)}_{j_2}t_2\right\}
			\cdots\exp\left\{-ia_{j_M}\hat{\mathfrak{h}}^{(n)}_{j_M}t_M\right\},
\end{equation}
which is a product of unitary evolutions,
for Hamiltonians~$\hat{\mathfrak{h}}^{(n)}_{j_1}$ represented by the sequence $j_i$,
and a sequence of times $t_i$.
The TS formul\ae~comprises a sequence of exponentials of Pauli operators that have
simple circuits for quantum computer implementation~\cite{NC00}.

Specifically, the approximation $U^{(n)}_\chi(\frac{t}{r})$ is constructed iteratively for the Hamiltonian 
	$H^{(n)}=\sum_{j=1}^m a_j \hat{\mathfrak{h}}^{(n)}_j$
via
\begin{align}
	{U}^{(n)}_{1}\left(\Delta t\right)
		=&\prod_{j=1}^m\exp\left\{-i a_j \hat{\mathfrak{h}}^{(n)}_j \frac{\Delta t}{2}\right\}
		\prod_{j=m}^1\exp\left\{-i a_j\hat{\mathfrak{h}}^{(n)}_j\frac{\Delta t}{2}\right\},\nonumber\\
	{U}^{(n)}_{p}\left(\Delta t\right)
		=&\left[U^{(n)}_{p-1}\left(\frac{s_pt}{r}\right)\right]^2U^{(n)}_{p-1}\left((1-4s_p)\frac{t}{r}\right)
			\left[U^{(n)}_{p-1}\left(\frac{s_pt}{r}\right)\right]^2
\label{eq:TSdef}
\end{align}
with
\begin{equation}
\label{eq:s_p}
	s_p=\frac{1}{4-4^{1/(2p-1)}}
\end{equation}
and the integer $p$ obeys $1<p\le \chi$.
We emphasize this form of the TS formula because it is important for our grouping algorithm that the order of the exponentials in the product formula matches the order of the exponentials in the Hamiltonian.

We express this TS stage of the algorithm as an outline of a computer program.
The program's input is the bit-string representation~$[\hat{H}^{(n)}]$ of the the $n$-qubit Hamiltonian,
the desired order of the Suzuki iteration~$\chi$,
the evolution time~$t$ and the number of intervals $r$, which together yield the time step~$\Delta t$ (\ref{eq:Deltat}).
The program's output bit-string representation for the $\chi^\text{th}$-order approximation to the true evolution:
\begin{equation}
 	\left[{U}^{(n)}_\chi(\Delta t)\right]
		\equiv (a_{j_M},{\bm l}_{j_M},\bS_{j_M},t_M)(a_{j_{M-1}},{\bm l}_{j_{M-1}},\bS_{j_{M-1}},t_{M-1})\cdots(a_{j_1},{\bm l}_{j_1},\bS_{j_1},t_1).\label{eq:encode}
\end{equation}
with $M=2m5^{\chi-1}$ following from the recursive form of the TS formul\ae~(\ref{eq:TSdef})~\cite{Suz90}.
The representation  in~(\ref{eq:encode}) stores each exponential in the approximation as a sequence of strings that represent a Hamiltonian term $a_{j}\mathfrak{h}^{(n)}_j$ and then the duration of the evolution step (stored to finite precision).  Our representation stores the exponentials in order of their execution; although, in this
case, the symmetry of the TS formul\ae~implies that we could also store them in reverse order without changing the result.
The TS algorithm that generates a TS approximation of the form~\eqref{eq:encode} is given by Algorithm~\ref{alg:LTS}.

\begin{algorithm}[t]
\caption{Trotter--Suzuki Algorithm.}
\label{alg:LTS}
\begin{algorithmic}
\Require \\
	$[\hat{H}^{(n)}]$: bit-string representation of the Hamiltonian\\
	$\Delta t$: time duration\\
	$\chi$: iteration order of Suzuki's method~\cite{Suz90}
\Ensure \\
	$\rm SuzInt$: array of exponentials.\\
\Function {TrotterSuzuki}{$[\hat{H}^{(n)}]$, $\Delta t$, $\chi$}
\State \Return ${\rm SuzInt} \gets$ sequence of exponentials $\left[{U}^{(n)}_\chi(\Delta t)\right]$~\eqref{eq:encode}
	using Suzuki's procedure~\cite{Suz90}.
\EndFunction
\end{algorithmic}
\end{algorithm}

The performance of the resulting simulation depends strongly on the chosen values for~$r$ and~$\chi$.
If $r=0$ or $\chi=0$, our program determines suitable values of $r$ or $\chi$;
otherwise this program uses the user-supplied values.
Given a specified value of~$\chi$,
then
\begin{equation}
\label{eq:rmin}
	r=\left\lceil 
		\frac{(2m(5/3)^{\chi-1}\chi\big(\max_j |a_j| t)\big)^{1+1/2\chi}}{(\epsilon/2)^{1/2\chi}}\right\rceil
\end{equation}
is optimal~\cite{WBHS10,PZ12},
which guarantees that~\cite{WBHS10}
\begin{equation}
\label{eq:Unt}
	\left\|U^{(n)}(t)-\left(\tilde{U}^{(n)}(\Delta t)\right)^r\right\|_2
		\le\frac{\epsilon}{2}
\end{equation}
provided that
\begin{equation}
	\epsilon \le {2m\chi(5/3)^{\chi-1}\max_j |a_j| t}.
\end{equation}
We employ the value of $r$ in~(\ref{eq:rmin}) as the default value of $r$ for the algorithm and take
our default value of $\chi$ to be
\begin{equation}
	\chi=\left\lceil \sqrt{\frac{\log_{25/3} (m \max_i |{a}_{i}| t/\epsilon)}{2}}\,\right\rceil
\end{equation}
because it causes the number of operations in the simulation to scale nearly linearly with $t$~\cite{WBHS10}.

The value of $r$ given in~(\ref{eq:rmin}) can be larger than necessary for certain Hamiltonians (as shown in~Section~\ref{appendix:numest}).
If a guarantee that the error is less than the tolerance~$\epsilon$ is not required,
then choosing~$r$ smaller than the optimal value given in~(\ref{eq:rmin}) could suffice and thereby reduce runtime.

\section{Hamiltonian Sorting Algorithm
\label{sec:grouping}}

Now we return to the first stage of the algorithm, which aims to reduce runtime by grouping TS terms based on 
generation by commuting Hamiltonians.
In other words, TS terms are grouped together to parallelize the quantum simulation circuit.
The benefits of grouping terms and exploiting parallelism is most notable in the case of physically local Hamiltonians, where we show that parallelism typically 
leads to a near--quadratic improvement to the scaling of the time required to simulate the quantum system's evolution.

The algorithm achieves this grouping by decomposing the  Hamiltonian~(\ref{eq:Hsum}) into $\bar{m}$ groups of terms as
\begin{equation}
\label{eq:GHsum}
	\hat{H}^{(n)}=\sum_{j=1}^{\bar{m}} \hat{\mathfrak{g}}_j^{(n)},
	\hat{\mathfrak{g}}_j^{(n)}=\sum_{l\in G_j} a_l\hat{\mathfrak{h}}_l^{(n)},
\end{equation}
such that all $\hat{\mathfrak{h}}_l^{(n)}\in G_j$ mutually commute.

\begin{algorithm}[t!]
\caption{Hamiltonian sorting algorithm.}
\label{alg:sorting}
\begin{algorithmic}
\Require \\
	$[\hat{H}^{(n)}]$: bit-string representation of the Hamiltonian\\
	$n$: number of qubits\\
	$m$: number of Hamiltonian terms summed to make $[\hat{H}^{(n)}]$
\Ensure \\
	$[\hat{H}_\text{sorted}^{(n)}]$: sorted representation of the Hamiltonian with mutually commuting terms combined in groups\\
\Function {SortH}{$[\hat{H}^{(n)}]$, $n$, $m$}
\State $m_{\rm s} \gets 1$.
\State $G_1 \gets (a_1, {\bm l}_1, {\bm S}_1)$.
\For{$j$ From $2$ to $m$}
\State ${\rm isAssigned} \gets 0$.
\For{$p$ from $1$ to $m_{\rm s}$}
\If{${\rm isAssigned}=0$} \Comment{Checks if term commutes with terms in $G_p$}
\State ${\rm isAssigned}\gets 1$ if $\sum_{v\ne w}\left|{\bS_v}_i \cap {\bS_w}_j\right|$ for each  $(a_i, {\bm l}_i, {\bm S}_i)$ in $G_p$.
\If{${\rm isAssigned}=1$}
\State $G_p \gets $ Concatenation of $G_p$ with $(a_j, {\bm l}_j, {\bm S}_j)$. \Comment{Assigns term to $G_p$.}
\EndIf
\EndIf
\EndFor
\If{$\rm isAssigned=0$}\Comment {Checks if new ``group'' of commuting Hamiltonians is needed}
\State $m_s\gets m_s+1$.
\State $G_{m_s} \gets (a_j, {\bm l}_j, {\bm S}_j)$.
\EndIf
\EndFor
\State \Return $[H_{\rm sorted}] \gets$ Concatenation of $G_1,\ldots,G_{m_s}$.
\EndFunction
\end{algorithmic}
\end{algorithm}

The Trotter--Suzuki algorithm can then be used to determine a sequence of exponentials of 
$\hat{\mathfrak{g}}_j^{(n)}$, namely the sum of the terms in each commuting set  that simulates 
\begin{equation}
	\exp{\left(-i\hat H^{(n)}t\right)}= \exp{\left(-i\sum_{j=1}^{\bar m}\hat{g}^{(n)}_j t\right)}
\end{equation}
within error $\epsilon$.  
Product-formula approximations are not needed
to decompose the exponential of each group into a product of exponentials of Pauli--operations
as each $\hat{\mathfrak{h}}^{(n)}_j$ in any given group mutually commutes.
In other words
\begin{equation}
	\exp{\left(-i\hat{\mathfrak{g}}_i^{(n)}t\right)} 
		= \exp{\left(-i\sum_{j\in G_i}{\hat{\mathfrak{h}}_j^{(n)}}t\right)}
		=\prod_{j\in G_i} \exp{\left(-i\hat{\mathfrak{h}}_j^{(n)}t\right)}
\label{eq:groupcommute}
\end{equation}
for any value of~$t$.

Finally, we note that explicitly grouping the terms in $\hat{H}^{(n)}$ into $\hat{\mathfrak{g}}^{(n)}$ is unnecessary.
Instead it suffices to sort the terms in the Hamiltonian by group membership
(i.e., terms that are assigned to group $G_1$ appear first in $[\hat{H}^{(n)}]$, then terms in $G_2$ appear next and so forth).
If we use~(\ref{eq:TSdef}) to construct the TS formul\ae,
then the resulting simulation will sort the steps in the simulation into commuting groups of operations.  
As commuting operations can often be executed in parallel,
this procedure reduces the time required to execute the circuit in systems that can use parallelism.

In the first stage of the algorithm our program accepts the string $\bS_i$,
as introduced in Sec.~\ref{sec:algorithm}, to determine if two terms in $\hat{H}^{(n)}$ commute. The Hamiltonians $\hat{\mathfrak{h}}_i^{(n)}$ and $\hat{\mathfrak{h}}_j^{(n)}$ commute if and only if,
for the dummy variables $v,w\in\{X,Y,Z\}$,
\begin{equation}
	\sum_{v\ne w}\left|{\bS_v}_i \cap {\bS_w}_j\right|\!\equiv\!  0\;({\rm mod}~ 2)
\label{eq:modeq},
\end{equation}
where $|\bullet|$ denotes the size of a set~$\bullet$.  

As a clarifying example, consider the Hamiltonian~(\ref{eq:Hn}). 
The first two terms in~(\ref{eq:Hn}) commute because of the anti--commutativity of Pauli--operators and because both terms have differing actions on an even number of qubits.  
Criterion~(\ref{eq:modeq}) also tells us that they commute because
\begin{equation}
	\sum_{v\ne w}|{\bS_v}_1 \cap {\bS_w}_2|\ =|{\bS_x}_1 \cap {\bS_y}_2|\ =2\equiv\!0~({\rm mod}~ 2).
\end{equation}
On the other hand, the second and third terms do not commute as
\begin{equation}
	\sum_{v\ne w}|{\bS_v}_2 \cap {\bS_w}_3|\ =|{\bS_y}_2 \cap {\bS_z}_3|\ =1\equiv\!1~({\rm mod}~ 2).
\end{equation}
Criterion~(\ref{eq:modeq}) accounts for the commutativity of Hamiltonians that act on disjoint sets of qubits as well
as other commuting Hamiltonians such as the star and plaquette operators in the toric code~\cite{Kit06}.  
A proof of the general validity of~(\ref{eq:modeq}) as a criterion for commutativity is given in~Appendix \ref{appendix:pauli}.
If desired, a more restrictive grouping condition
\begin{equation}
	\sum_{v\ne w}|{\bS_v}_i \cap {\bS_w}_j|\!\equiv\!  0\label{eq:grouprestrict}
\end{equation}
can be used only to group together operations that have actions on disjoint sets of qubits.

The criterion for commutativity in~(\ref{eq:modeq}) can be used to find an efficient classical algorithm for grouping  
$\{\hat{\mathfrak{h}}_j^{(n)}\}$ into groups of mutually commuting Hamiltonians, which we describe below formally.
This grouping algorithm is efficient because $m$ is polynomially large in $n$ for local--Hamiltonians 
and~(\ref{eq:modeq}) can be efficiently evaluated.

The depth of the resulting quantum circuit depends on the value of $\bar m$ for the Hamiltonian.  The 
reductions in the depth vary with the number of groups required for the Hamiltonian
in question.  The question can, however, be addressed for the cases of generic $k$--local or physically $k$--local Hamiltonians (by generic we mean $k$-local Hamiltonians that include every possible $p$-body interaction for $p\le k$.).

We estimate the worst--case scaling of $\bar m$ for generic $k$-local $\hat{H}^{(n)}$ by using the more restrictive grouping condition that $\hat{\mathfrak{h}}^{(n)}_i$ and $\hat{\mathfrak{h}}^{(n)}_j$ are assigned to the same group only if
condition~(\ref{eq:grouprestrict}) is satisfied.
This requirement will generically  result in a smaller value of $\bar m$ than our grouping algorithm will yield.  In this case, at most $\lfloor n/k \rfloor$ $k$-body terms can be assigned to each group for $k$--local interactions.  Terms with $(k-1)$--body interactions (or fewer) can be neglected because $n$ is assumed to be large and the vast majority of terms in the $k$--local Hamiltonian are $k$--body.  
Specifically, there are $O(n^{k-1})$ terms with only $(k-1)$--body interactions or fewer and $O(n^k)$ terms with $k$--body interactions,
which means that
we can neglect terms that are only $(k-1)$--local for generic Hamiltonians in the limit of large $n$.  

The number of $k$--body Hamiltonians that can be assigned to each group before the Hamiltonians violate the grouping criterion~(\ref{eq:modeq}), scales as $\Theta(n/k)$,
with $\Theta$ the Bachmann-Landau notation for indicating that a function of this order is asymptotically bounded above and below by $n/k$ up to multiplicative constants.
For~$k$ a constant
\begin{equation}
\label{mbar}
	\bar{m} \in O\left( \frac{n^{k}}{n/k}\right) = O(n^{k-1}).
\end{equation}
Physically local Hamiltonians are constrained such that each qubit interacts with at most a constant number of qubits.
This implies that there are $O(n)$ $k$--body terms present in physically $k$--local Hamiltonians.  
Therefore, the number of groups required for physically--local Hamiltonians scales as
\begin{equation}
\label{mbar2}
	\bar{m} \in O\left( \frac{n}{n/k}\right) = O(1),
\end{equation}
 which we 
will see constitutes a nearly--quadratic reduction in the circuit depth for some simulations.

\section{Implementing Trotter--Suzuki Formulas~\label{sec:implement}}

The main primitive element of the circuit-design algorithm is a basic circuit element $\mathcal{C}_\ell$
corresponding to a particular sequence of gates in our gate set to simulate evolution due to each
$\exp\{-i a_{j_\ell} \hat{\mathfrak{h}}^{(n)}_{j_\ell} t_\ell\}$ 
in the output of the subroutine described in Sec.~\ref{sec:TS}.  The circuit construction that
we use is an optimized version of that presented in~\cite{NC00}.
Underpinning this primitive~$\mathcal{C}_\ell$ is the conversion of Pauli gate operations to operations in our gate set~\cite{WML+10,WMBL11,WBA11}.
\begin{figure}
\includegraphics[width=0.75\textwidth]{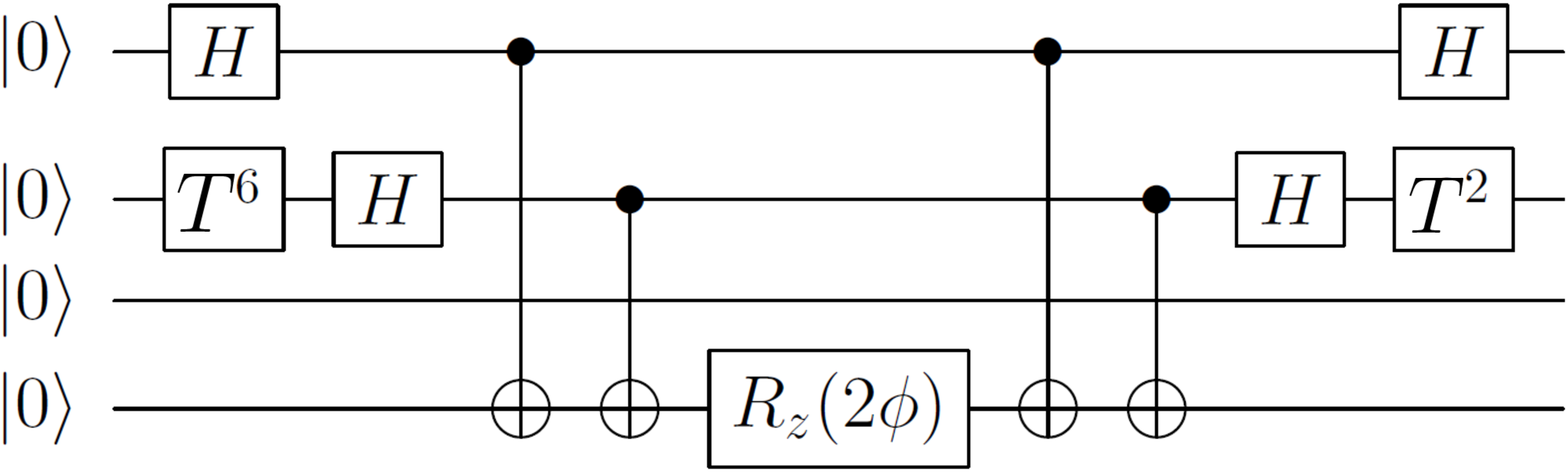}
\caption{Quantum circuit for implementing $\exp(-i\phi X\otimes Y\otimes\openone\otimes Z)$ for an arbitrary dimensionless evolution time $\phi$ with $H$ the Hadamard gate, $T^\ell$ a concatenation of $\ell$ $\pi/8$ gates, and $R_Z(2\phi)$ a qubit rotation of $\phi$ about $Z$.}
\label{fig:circuit}
\end{figure}

The simplest way to explain this step is by example:
\begin{equation}
\label{eq:example}
	U_\ell^{(n=4)}(\phi)=\exp\left(-i\phi X\otimes Y\otimes\openone\otimes Z\right)
\end{equation}
shown in Fig.~\ref{fig:circuit}.
This circuit allows for a continuously parametrized phase-rotation gate $R_z(2\phi)=\exp(-i\phi Z)$,
but of course a proper quantum algorithm would work with a finite gate set.
However, the gate $R_Z(2\phi)$ can be reduced to a finite gate set using the constructive version of the Solovay-Kitaev algorithm~\cite{DN06}.
Our algorithm takes a logical variable,$\varpi\in\{0,1\}$ as an input that specifies the gate set from which the
gates in the output should be drawn and uses the Dawson--Nielsen algorithm~\cite{DN06} to convert the
continuous rotation gates into discrete gates if gate set $0$ is chosen. We label the gate sets
\begin{itemize}
\item[] $\varpi=0$: $\{{\rm H,\Tgate,CNOT}\}$,
\item[] $\varpi=1$: $\{{\rm H}, R_z(\theta), \text{CNOT}\colon \theta\in [0,2\pi)\}$.
\end{itemize}




The circuit primitive for $\mathcal{C}_\ell$, as illustrated in Fig.~\ref{fig:circuit}, is constructed by first choosing
one of the system qubits to be the `parity qubit'. The role of this qubit is to track the parity of qubits affected
by $\hat{\mathfrak{h}}_{j_\ell}^{(n)}$ when expressed in the eigenbasis of $\hat{\mathfrak{h}}_{j_\ell}^{(n)}$.
This parity is required to perform $U_{\ell}^{(n)}(a_{j_\ell} t_\ell)$ using diagonalization~\cite{WML+10,WMBL11,WBA11}.
The parity qubit is always chosen to be the qubit with the largest label amongst this set that is non-trivially affected by the Hamiltonian. We say that a qubit is trivially affected by a Hamiltonian if it acts as the identity on that qubit.  
As an example, the fourth (bottom) qubit is chosen to be the parity qubit for the Hamiltonian in Fig.~\ref{fig:circuit}.

The method we use to construct the simulation circuit is given in Algorithm~\ref{alg:pauli}.
Specifically this stage of the algorithm produces a bit-string representation for a circuit approximating
$\exp\{-ia_i\mathfrak{h}_i^{(n)} t_i\}$ where $a_i\mathfrak{h}_i^{(n)}$ is a term in $\hat{H}^{(n)}$.  
  The algorithm for this is provided below. 
 
\begin{algorithm}[t]
\caption{Circuit-design algorithm for Pauli--exponentials.}
\label{alg:pauli}
\begin{algorithmic}
\Require \\
	$({ a}_i, {\bm l}_i, {\bm S}_i, t_i)$: bit-string representation of the exponential\\
	$\varpi$: determines which gate set should be used\\
	$\delta$: error--tolerance for the Solovay--Kitaev Algorithm
\Ensure \\
	$[\mathcal{C}_i]$: simulation circuit for $\exp({-i \hat{\mathfrak{h}}^{(n)}_{i} t_i})$
\Function {PCircuit}{$({\bm a}_i, {\bm l}_i, {\bm S}_i, t_i)$,$\varpi$, $\delta$}
\State $[\mathcal{C}_i] \gets \emptyset$.\Comment{Sets $[\mathcal{C}_i]$ to the empty--string.}
\For{each string $\ell$ in ${\bm S}_{i,x}$}
\State $[\mathcal{C}_i]\gets[\mathcal{C}_i]H\ell$.\Comment{Concatenates $[\mathcal{C}_i]$ with Hadamard on qubits that $\hat{\mathfrak{h}}^{(n)}_i$ acts as $X$ on.}
\EndFor
\For{each string $\ell$ in ${\bm S}_{i,y}$}
\State $[\mathcal{C}_i]\gets[\mathcal{C}_i]\Tgate\ell \Tgate\ell\Tgate\ell \Tgate\ell\Tgate\ell \Tgate\ell H\ell$.\Comment{Applies diagonalizing rotation to each qubit on which the term acts as $Y$.}
\EndFor
\State $\ell_{\max}\gets \max(\ell \in {\bm S}_i)$.
\For{each $\ell$ in ${\bm S}_i \setminus \{\ell_{\max}\}$}\Comment{Identifies parity qubit.}
\State $[\mathcal{C}_i]\gets [\mathcal{C}_i]{\rm CNOT}\ell,\ell_{\max}$.
\EndFor
\If{$\varpi$=1}
\State $SK\gets$ output of the Solovay--Kitaev Algorithm for $R_z(2{ a}_i t_i)$ acting on qubit $\ell_{\max}$ with error tolerance $\delta$.
\State $[\mathcal{C}_i]\gets [\mathcal{C}_i]SK$.
\Else
\State $[\mathcal{C}_i]\gets [\mathcal{C}_i]{\rm RZ}(2a_it_i),\ell_{\max}$.
\EndIf
\For{each $\ell$ in ${\bm S}_i \setminus \{\ell_{\max}\}$}
\State $[\mathcal{C}_i]\gets [\mathcal{C}_i]{\rm CNOT}\ell,\ell_{\max}$.
\EndFor
\For{each string $\ell$ in ${\bm S}_{i,y}$}
\State $[\mathcal{C}_i]\gets[\mathcal{C}_i]H\ell\Tgate\ell \Tgate\ell$.
\EndFor
\For{each string $\ell$ in ${\bm S}_{i,x}$}
\State $[\mathcal{C}_i]\gets[\mathcal{C}_i]H\ell$.
\EndFor
\State \Return $[\mathcal{C}_i]$.
\EndFunction
\end{algorithmic}
\end{algorithm}

\section{Main Algorithm}
\label{sec:mainalg}

\subsection{Complete Procedure}
\label{subsec:complete}
We now construct the circuit-design algorithm, which employs the programs described in the previous three sections.
The algorithm begins by using Algorithm~\ref{alg:sorting} to sort the terms in the Hamiltonian, which ensures that neighboring entries in the list of terms that comprise the Hamiltonian commute (if possible).  The next step uses the Trotter--Suzuki algorithm
to find a sequence of simulations of the $\hat{\mathfrak{h}}_{j_\ell}^{(n)}$ that approximates $e^{-i\hat{H}^{(n)} t}$.  The final step utilizes Algorithm~\ref{alg:pauli} to find a quantum circuit that approximates each of the $\exp\{-i\hat{\mathfrak{h}}_{j_\ell}^{(n)} t_\ell\}$ to yield a complete description of the overall simulation circuit.  The procedure is described in greater detail in Algorithm~\ref{alg:main}.

\begin{algorithm}
\caption{Main Algorithm}
\label{alg:main}
\begin{algorithmic}

\Require \\
	$[\hat{H}^{(n)}]$: bit-string representation of the Hamiltonian\\
	$n$: The number of qubits\\
	$t$: evolution time\\
	$\epsilon$: error tolerance\\
	$r$: number of time steps	\Comment{$r=0$ guarantees error is at most $\epsilon$}\\
	$\chi$: iteration order of TS formula	\Comment{$\chi=0$ guarantees near--linear time scaling.}\\
	$\varpi$: logical value that indicates whether the discrete or continuous gate setis used
\Ensure \\
	$[\mathcal{C}]$: simulation circuit for $\exp({-i \hat{H}^{(n)}_{i} t})$
\Function {Main}{$[\hat{H}^{(n)}],n, t,r,\chi,\epsilon,\varpi$}
\State $[\mathcal{C_{{\rm temp}}}]\gets \emptyset$.
\State $m\gets$  number of terms in $\hat{H}^{(n)}$.
\State $[\hat{H}^{(n)}]\gets {\rm SortH}([\hat{H}^{(n)}],n,m)$.\Comment{See Alg.~\ref{alg:sorting}}
\State $a_{\max}\gets \max_i |a_i|$.
\If{$\chi=0$}
\State $\chi\gets \left\lceil \sqrt{\frac{\log_{25/3}\left (\frac{m{ a}_{\max} t}{\epsilon}\right)}{2}}\,\right\rceil$.\Comment{Computes default value of $\chi$}
\EndIf
\If{$r=0$}
\State $r\gets \left \lceil\frac{2(2m(5/3)^{\chi-1}\chi { a}_{\max} t)^{1+1/(2\chi)}}{(\epsilon/2)^{1/(2\chi)}} \right \rceil$.\Comment{Computes default value of $r$}
\EndIf
\If{$\epsilon > {2m\chi(5/3)^{\chi-1}\max_j |a_j| t}$}
\State $\epsilon\gets {2m\chi(5/3)^{\chi-1}\max_j |a_j| t}$.
\EndIf
\State ${\rm SuzInt}\gets $ TrotterSuzuki$([\hat{H}^{(n)}],\frac{t}{r},\chi)$.\Comment{See Alg.~\ref{alg:LTS}}
\For{$j=1$ to $2m5^{\chi-1}$}\Comment{Finds circuit for one time step.}
\State $[\mathcal{C_{{\rm temp}}}]\gets [\mathcal{C_{{\rm temp}}}]{\rm PCircuit}\left({\rm SuzInt}(j),\varpi,\frac{\epsilon}{(4m5^{\chi-1}r)}\right)$.\Comment{See Alg.~\ref{alg:pauli}}
\EndFor
\State $[\mathcal{C}]\gets [\mathcal{C}_{\rm temp}]$.
\For{$j=1$ to $r-1$}\Comment{Finds complete simulation circuit.}
\State $[\mathcal{C}] \gets [\mathcal{C}][\mathcal{C}_{\rm temp}]$.
\EndFor
\State \Return $[\mathcal{C}]$.
\EndFunction
\end{algorithmic}
\end{algorithm}

The efficiency of our main algorithm depends on whether $r$ and $2m5^{\chi_0-1}$ are polynomially large.  It is straightforward to see by substitution that
the default values used for both of these quantities scale polynomially with the simulation parameters, and hence is efficient.  On the other hand, if the
default values of $r$ and $\chi$ are not used then the above algorithm may not be efficient.

\subsection{Cost Estimates \label{sec:costs}}
We now provide upper bounds for the scaling of the circuit--size of the circuits yielded by our design algorithm using the
default values $\chi_0$ and $r_0$, which are chosen to guarantee that the simulation time scales near--linearly with $t$ and
the error is at most $\epsilon/2$ respectively.  There are three costs that we consider: the number of gates in the resulting
circuit, $N_{\rm op}$, the time required to execute the circuit using parallelism, $\tau$, and the number of qubits required which
in our case is trivially $n$.  We assess the remaining costs $N_{\rm op}$ and $\tau$ below.

The value of $N_{\rm op}$ is bounded above by the number of circuit primitives, $\mathcal{C}_\ell$, needed to simulate
the evolution multiplied by the maximum cost of implementing a circuit primitive.
{The cost for implementing each circuit primitive~$\mathcal{C}_\ell$ for a $k$-local $\hat{H}$ requires at most $10k$ single-qubit gates,
$2k-2$ CNOT gates and one $R_Z$ rotation.  This worst--case estimate is given by the cost of simulating a $\hat{\mathfrak{h}}_j^{(n)}$ that acts as $Y$ on $k$ qubits because $Y$--interactions are the most expensive to simulate
using our algorithm for finding $\mathcal{C}_\ell$.
The Solovay--Kitaev algorithm of Dawson and Nielsen~\cite{DN06} gives the cost of implementing
the continuous gate $R_Z(2\phi)$ within tolerance 
$\epsilon/(4m5^{\chi_0-1} r_{0})$ as
\begin{equation}
N_{\rm SK}\in O\left(\log^{4}\left(\frac{4m5^{\chi_0-1}(2m(5/3)^{\chi_0-1}\chi { a}_{\max} t)^{1+1/(2\chi_0)}}{2\epsilon(\epsilon/2)^{1/(2\chi_0)}}\right)\right).
\end{equation}
As 
\begin{equation}
\chi_0=\left\lceil \sqrt{\frac{\log_{25/3} (m{ a}_{\max} t/\epsilon)}{2}}\,\right\rceil,
\end{equation}
we have that $\log(5^{\chi_0})\in O(\sqrt{\log(m a_{\max} t/\epsilon)})$.  Then using the properties of logarithms, we find that
\begin{equation}
\label{eq:N_phi}
	N_{\rm SK}\in O\left(\log^4\left(\frac{m\max_i |a_i| t}{\epsilon}\right)\right).
\end{equation}
The total number of operations used to implement each primitive circuit $\mathcal C_\ell$ within tolerance~$\epsilon$ is thus $O(k+N_{\rm SK})$.
As $\mathcal{C}=\mathcal{C}_M\cdots\mathcal{C}_1$, the total number of operations in $\mathcal{C}$ scales as $O\big( M(k+N_{\rm SK})\big)$.
Finally, $U^{(n)}(t)$ is simulated by $\mathcal{C}^r$ so the total number of operations scales as 
\begin{equation}
N_{\rm{op}}\in O\left( 2m5^{\chi_0-1}r_{0}(k+N_{\rm SK})\right).
\end{equation}  
We then substitute the value of $r_0$ into this expression to find that,
\begin{equation}
	N_{\rm op}
		\in\frac{(k+N_{\rm SK})m^{2+o(1)}\left(\max_i |a_i|t\right)^{1+o(1)}}{\epsilon^{o(1)}},
\label{eq:Nbdgenb}
\end{equation}
with $\epsilon^{o(1)}$, $m^{o(1)}$, and $(\max_i |a_i|t)^{o(1)}$ representing quantities that scale sub-polynomially but not quite poly-logarithmically.
As $N_{\rm SK}$ varies sub-polynomially with respect to all parameters and $k$ is a constant,
$N_{\rm SK}+k$ can be incorporated into $\left(m\max_i |a_i|t/\epsilon\right)^{o(1)}$ in Relation~(\ref{eq:Nbdgenb}).  This leads to the conclusion that
\begin{equation}
	N_{\rm op}
		\in\frac{(k+N_{\rm SK})m^{2+o(1)}\left(\max_i |a_i|t\right)^{1+o(1)}}{\epsilon^{o(1)}}\in \frac{m^{2+o(1)}\left(\max_i |a_i|t\right)^{1+o(1)}}{\epsilon^{o(1)}}.
\label{eq:Nbdgenb2}
\end{equation}

The performance of our circuit-design algorithm is enhanced if a reasonable extra restriction is placed on the $k$-local $\hat{H}$.
The upper bound~$m$ is different for $k$-local vs physically $k$-local $\hat{H}$ as we now see.
For $k$-local $\hat{H}^{(n)}$,
\begin{equation}
	m\le \sum_{q=1}^k 3^{q}\binom{n}{q}\in O( n^k),
\end{equation}
because there are at most $3^{q}\binom{n}{q}$ $q$-body terms in $\hat{H}^{(n)}$ for $q=1,\dots,k$.
The scaling $m\in O(n^k)$ arises from standard inequalities for binomial sums and, as~$k$ is a constant, then so is~$3^k$.
If $\hat{H}^{(n)}$ is physically $k$-local, $m\in O(n)$ because each qubit interacts with at most a constant number of neighbors.

We can then eliminate $m$ from~(\ref{eq:Nbdgenb}) by noting that, if $\hat{H}^{(n)}$ is $k$-local, then $m\in O(n^k)$ for $k$ a constant, and
\begin{equation}
	N_{\rm op}\in \frac{n^{k(2+o(1))}\left(\max_i |a_i|t\right)^{1+o(1)}}{\epsilon^{ o(1)}}.
\label{eq:UB}
\end{equation}
If $\hat{H}^{(n)}$ is physically $k$-local and $k$ is constant,
then $m\in O(n)$.  We substitute this scaling
into~(\ref{eq:Nbdgenb}) to obtain
\begin{equation}
	N_{\rm op}\in \frac{n^{2+o(1)}\left(\max_i |a_i|t\right)^{1+o(1)}}{\epsilon^{o(1)}}.
\label{eq:UB_scaling_for_PLH}
\end{equation}
Comparing (\ref{eq:UB}) to (\ref{eq:UB_scaling_for_PLH}) shows that the simulation cost  is dramatically reduced for $\hat{H}^{(n)}$ physically $k$-local rather than just $k$-local.  This cost reduction does not occur from a modification of the algorithm, but rather a more careful costing of the performance of our algorithm.

In fact the circuits generated by our algorithm are optimal, or near-optimal, in three distinct ways.  First, they exhibit near-optimal scaling with $t$ because linear scaling is known to be a lower bound
for general quantum simulation~\cite{BACS07a,BACS07b,CK10}.  Second, they have optimal space complexity because a minimum of $n$-qubits of memory is needed to simulate the quantum dynamics of an $n$ qubit system.
Finally, the $n$-scaling of~(\eqref{eq:UB}) and~(\ref{eq:UB_scaling_for_PLH}) is unlikely to be surpassed
by other general purpose TS-based simulation algorithms because the scaling with~$n$ is derived from the value of~$m$ for the Hamiltonian~\cite{BACS07a,BACS07b}.

The fact that better scaling with $n$ cannot be obtained by using a superior decomposition method for the Hamiltonian follows from  Vizing's edge-coloring graph algorithm~\cite{Viz64},
which states that a graph with maximum degree $d$ cannot be colored using fewer than $d$ colors.  This implies that a $d$--sparse Hamiltonian can be decomposed into at best $d$ one-sparse matrices.  A $k$--local Hamiltonian can be at most $O(n^k)$ sparse, which implies that $O(n^k)$ terms will be present in the Hamiltonian using the optimal decomposition method.  
This value of~$m$ coincides with the value that our algorithm finds for simulating $k$-local Hamiltonians;
hence our algorithm is unlikely to be significantly surpassed by other algorithms that use similar strategies.

Now we will examine the scaling of the time required to implement the resultant quantum circuits on quantum computers that can exploit parallelism.
Without grouping, the depth of the quantum circuits yielded by our algorithm scales with $m$ is at worst $m^{2+o(1)}$.  Although our grouping step does not change the circuit size, it causes the depth of the resulting
circuits to scale as $\bar{m} m^{1+o(1)}$, where $\bar m$ may be smaller than $m$.

The factor of $m^{1+o(1)}$ remaining in the scaling comes from the upper--bound used to estimate error in
the Trotter--Suzuki formulas, which does not change if grouping is used.
Thus, parallel execution of the exponents in each group can be used to reduce the scaling of the execution time of the quantum simulation, $\tau$, for $k$-local Hamiltonians to
\begin{equation}
\tau\in \frac{n^{k(2+o(1))-1}\left(\max_i |a_i|t\right)^{1+o(1)}}{\epsilon^{ o(1)}},
\end{equation} 
and for the case of physically $k$-local Hamiltonians it becomes 
\begin{equation}
\tau \in \frac{n^{1+o(1)}\left(\max_i |a_i|t\right)^{1+o(1)}}{\epsilon^{o(1)}},
\end{equation}
which scales nearly--quadratically better with  $n$ than what we would expect if grouping were not used.


\section{Examples}
We now examine the performance of our circuit construction algorithm when applied to simulating the quantum dynamics of two important physical systems.  
Specifically, we examine simulating Kitaev's Honeycomb model and pairing models similar to the Bardeen--Cooper--Schreifer model of superconductivity.

\subsection{Simulating Kitaev's Honeycomb Model}
Consider Kitaev's honeycomb model~\cite{Kit06} described by the Hamiltonian
\begin{equation*}
	\hat{H}^{(n)} = -J_{x}\sum_{x-{\rm link}}X_iX_j-J_{y}\sum_{y-{\rm link}}Y_iY_j-J_{z}\sum_{z-{\rm link}}Z_iZ_j,
\end{equation*}
with the links shown in the honeycomb-lattice representation depicted in Fig.~\ref{fig:honeycomb}.
\begin{figure}\centering
\includegraphics[width=0.3\linewidth]{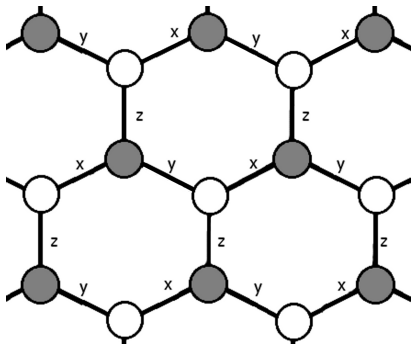}
\caption{
	Kitaev's honeycomb lattice with each vertex representing a physical qubit and each edge representing an interaction between two qubits
	denoted by~$x$-, $y$-, and $z$-links.
\label{fig:honeycomb}\protect}
\end{figure}
Although exactly solvable~\cite{Kit06}, the ground state has applications for topological error correction and is difficult to experimentally prepare.
Our simulation circuits can then be used (in conjunction with a ground-state preparation method such as adiabatic state preparation~\cite{WBL02} or the Abrams Lloyd algorithm~\cite{AL97}) to prepare the ground states of such Hamiltonians.

The resulting sequence of exponentials yielded by our circuit design algorithm yields a sequence of exponentials of $X_i X_j$, $Y_i Y_j$ and $Z_i Z_j$.  Figure~\ref{fig:simcirc} gives the simulation circuits that our algorithm uses to simulate each of these exponentials.  We can use these diagrams to find the number of operations used in the simulation by using the fact that each term in the Hamiltonian appears $2(5)^{\chi-1}$ times in the TS formula and that there are $n/2$ different $XX$, $YY$ and $ZZ$  interaction terms in the Hamiltonian.  

As~$r$ TS formulas are used in the simulation, the total number of times each of these three types of interactions appears is $n5^{\chi-1}r$.  The total number of gates required for the simulation can be found by multiplying the number of interactions of each type by the number of gates needed to simulate that type of interactions (explicit constructions for these circuits are given in Fig.~\ref{fig:simcirc}).  The total number of operations required to simulate the Honeycomb model is summarized below.

\begin{itemize}
\item Hadamard gates: $8n5^{\chi-1} r$,
\item \Tgate~gates: $16n5^{\chi-1} r$,
\item Z--Rotation gates: $3n5^{\chi-1} r$,
\item C-NOT gates: $6n5^{\chi-1} r$,
\end{itemize}
where $r$ is the number of time steps used in the simulation and $\chi$ is the iteration order of the Trotter--Suzuki formula
used in the simulation.  If $r$ is chosen as per~(\ref{eq:rmin})
then the error is promised to be less than $\epsilon/2$ given $\epsilon$ is sufficiently small~\cite{BACS07a,BACS07b}; however, other choices of $r$ are possible if rigorous guarantees that the simulation error is less than $\epsilon$ are not desired.

We can then directly estimate the scaling of the circuit size with the simulation parameters if $\chi$ and $r$ are taken to be the default values for our 
algorithm.
Kitaev's honeycomb-model Hamiltonian is physically two-local with $m \le 3n/2$.
Combining the observation that \begin{equation}\max_i |a_i|\le \max\left\{|J_x|,|J_y|,|J_z|\right\},\end{equation} with~(\ref{eq:UB_scaling_for_PLH})
implies that our circuit-design algorithm yields $\mathcal C$ for simulating $U^{(n)}(t)$ 
within error tolerance $\epsilon$ and with a circuit size that scales as
\begin{equation}
	N_{\rm op}\in n^{2+o(1)}\left(\max\{|J_x|,|J_y|,|J_z|\}t\right)^{1+o(1)}/\epsilon^{o(1)},
\label{eq:UB_scaling_for_HC}
\end{equation}	
elementary gates from~$\mathcal G$ acting on only $n$ qubits.
This scaling is significantly better than previous algorithms, which had a bound on the number of gates in $O(n^4\log^* n)$
and utilize quantum oracles that may be difficult to implement~\cite{BACS07a,BACS07b}.
\begin{figure}[t!]
\centering
\includegraphics[width=0.7\linewidth]{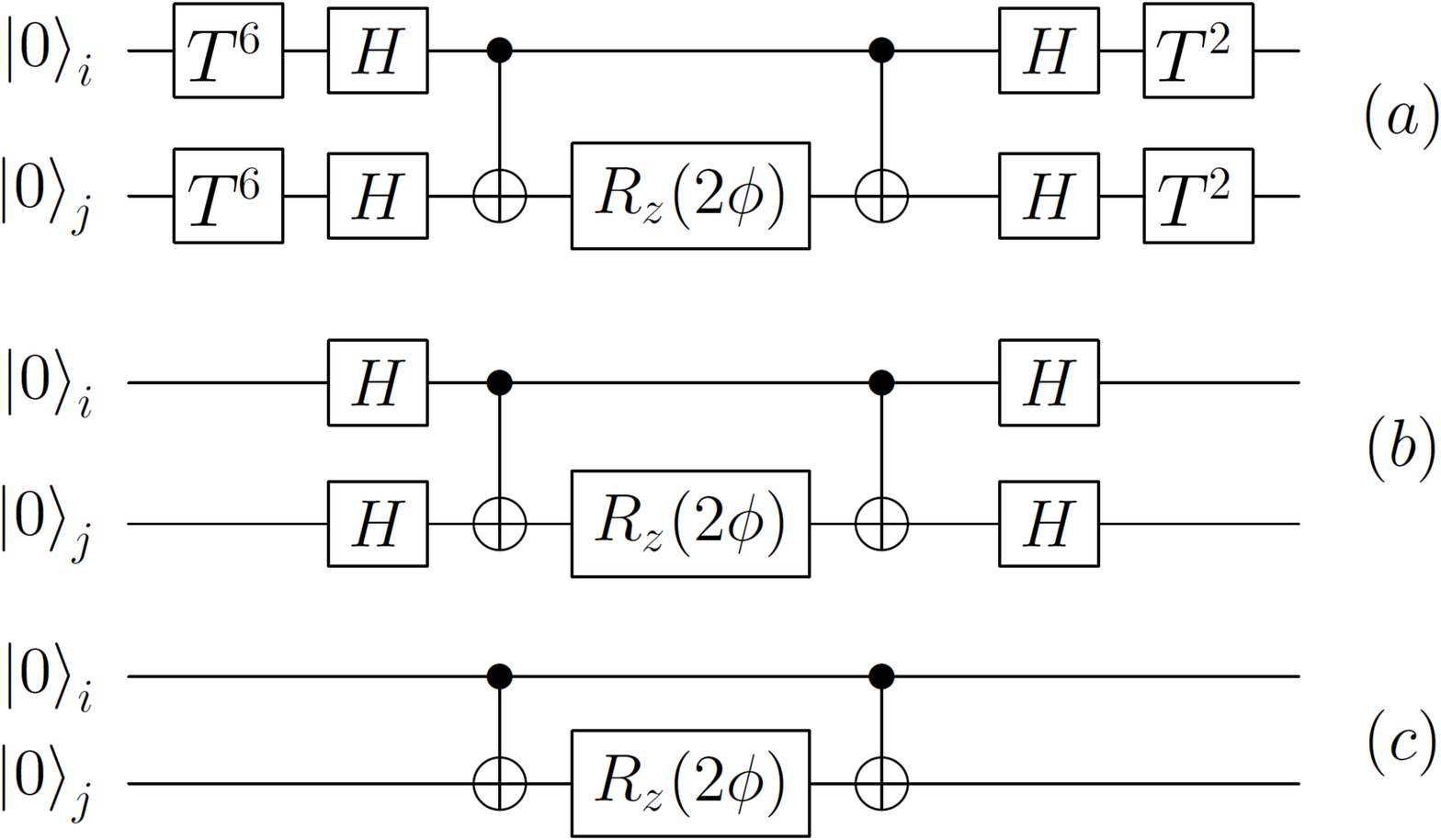}
\caption{Circuits for simulating evolution due to exponentials of the terms present in the honeycomb model or pairing models: (a) simulates $e^{-i Y_iY_j \phi}$ (b) simulates $e^{-i X_iX_j \phi}$ and (c) simulates $e^{-i Z_iZ_j \phi}$.
Exponentials of the form $e^{-i Z_p \phi}$ can be trivially simulated with an $R_z$ rotation.\label{fig:simcirc}}
\end{figure}

\subsection{Simulating Pairing Models}

Our second example simulates general pairing Hamiltonian evolution, which is central to studies of superconductivity in many-body systems.
A notable example of such Hamiltonians is the BCS Hamiltonian, which describes the interaction of electrons on a lattice according to~\cite{RS80,Mah00}
\begin{equation}
\label{eq:BCS}
	\hat{H}_{\rm BCS}=\frac{1}{2}\sum_{p=1}^n\mathcal{E}_p(a_{p}^\dagger a_p+ a_{-p}^\dagger a_{-p})
		+\sum_{p,l=1}^n V_{pl} \hat{a}_p^\dagger \hat{a}_{-p}^{\dagger}\hat{a}_l^{~} \hat{a}_{-l}^{~},
\end{equation}
where $\hat{a}_p$ is the fermionic annihilation operator for a fermion in states~$p=(\bm{p},\uparrow)$
and $-p=(-\bm{p},\downarrow)$ with~$\bm{p}$
the particle's momentum and $\uparrow$ and $\downarrow$ its spin,
$\hat{N}_p$ is the number operator for fermions in state $p$,
$\mathcal{E}_p$ is the on-site interaction strength,
and $V_{pl}$ is the interaction strength between neighboring fermions.

The general BCS Hamiltonian~(\ref{eq:BCS}) can be mapped to a spin system by using each spin to represent the presence or absence of a fermion in that mode.
Wu, Byrd and Lidar~\cite{WBL02} show that a large class of pairing models that subsume the BCS Hamiltonian can be expressed as
\begin{equation*}
	H_p=\frac{1}{2}\sum_{p=1}^n \gamma_pZ_p +\sum_{r=\pm}\sum_{l>p=1}^n V_{pl}^r\left(X_pX_l +rY_pY_l\right),
\end{equation*}
for $X_p$, $Y_p$ and $Z_p$ the Pauli $X$, $Y$ and $Z$ operators applied to qubit $p$.  Specifically, the BCS Hamiltonian has $V_{pl}^- =0$ and $\gamma_p=\mathcal{E}_p+V_{pp}$.

As in the previous example, our circuit design algorithm yields a sequence of exponentials of the different terms in the Hamiltonian.  The circuit implementations of the resulting exponentials can also be seen in Fig.~\ref{fig:simcirc}.  By multiplying the total number of exponentials of each type by the number of gates used in the implementation of each exponential and collecting the result, we find that our algorithm yields a circuit containing the following numbers of gates:
\begin{itemize}
\item Hadamard gates: $16n(n-1)5^{\chi-1} r$,
\item \Tgate~gates: $32n(n-1)5^{\chi-1} r$,
\item Z-- Rotation gates: $2n(2n-1)5^{\chi-1} r$,
\item C-NOT gates: $8n(n-1)5^{\chi-1} r$,
\end{itemize}
where the algorithms cited scaling is found by choosing $r$ as per
Eq.~(\ref{eq:rmin}) and substituting the asymptotic scaling of $r$ for $\chi$ chosen approximately optimally~\cite{BACS07a,BACS07b}. The value of $\chi$ that reduces the \emph{total number} of gates can be found by minimizing the sum of these gate counts over all $\chi$.

Previously,
circuit design yielded circuits with $O(n^5 t^2)$ gates with no promises about the accuracy of the simulation~\cite{WBL02}.  The circuits yielded by our algorithm are polynomially shorter than this previous best method.  This follows from the fact that
$H_p$ is two-local and hence $m\in O(n^2)$,
which implies that our algorithm yields a circuit with complexity
\begin{equation}
	N_{\rm op}\in \frac{n^{4+o(1)}\left(\max_{p,l,r}\{|\gamma_p|, |V_{pl}^r|\}t\right)^{1+o(1)}}{\epsilon^{o(1)}},\label{eq:wblscale}
\end{equation}
when the default values of $r$ and $\chi$ are used.

\begin{figure}[t]
\centering
\includegraphics[width=0.8\textwidth]{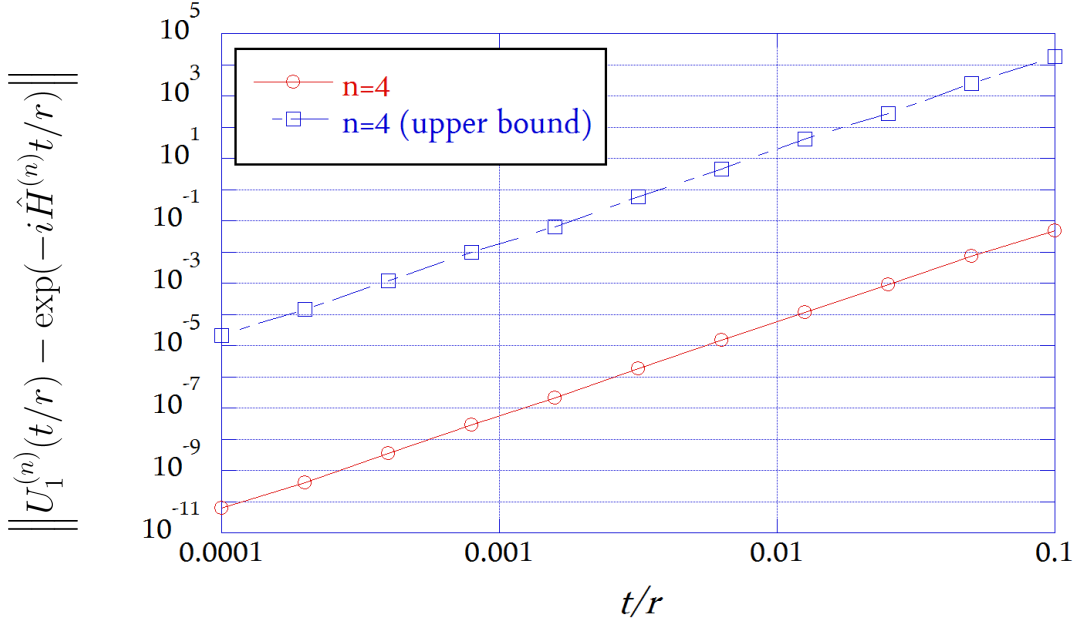}
\caption{
This plot shows the mean error or mean upper bound for the error incurred when using the lowest--order Trotter--Suzuki formula on $50$ randomly generated two-body Hamiltonians acting on $4$ qubits that were sampled from our Gaussian ensemble of Hamiltonians.  This result shows that the lowest--order Trotter--Suzuki formula given in~(\ref{eq:ltsbd}) can be too loose by as much as six orders of magnitude when applied to simulations of $2$--local Hamiltonians.  
\label{fig:cmperror}\protect}
\end{figure}

Our algorithm can thus generate efficient quantum circuits for simulating the dynamics of general BCS Hamiltonians.
Our resultant circuits can be used in conjunction with eigenvector simulation techniques or adiabatic state preparation to approximate the ground state of a quantum system
and can thus be useful for autonomously determining whether classes of pairing models afford exotic types of superconductivity.
\section{Numerical Estimates of Error in Low--Order Trotter--Suzuki Formul\ae~for 2--Body Hamiltonians\label{appendix:numest}}
The results of the previous section suggest that quantum simulations of pairing Hamiltonians may be difficult for large $n$ because the number of required operations scales, at most, as $n^{4+o(1)}$; however, we do not know whether this upperbound  is tight.
  Here we provide numerical evidence that the error bounds on which this scaling is based can substantially overestimate the error in the Trotter--Suzuki formula for a given simulation.  The upper bound in question is used to
estimate the number of time steps, $r$, needed in the simulation and therefore we will conclude that the  upper bounds for $r$ cited in~\cite{BACS07a,BACS07b,WBHS10}
are too loose for some practical cases.  


We analyse the Trotter--Suzuki error for Hamiltonians chosen randomly from an ensemble of $2$-body Hamiltonians that have their $a_j$ independently distributed according to a Gaussian with mean zero and unit variance.  We choose two--body, rather than $2$--local Hamiltonians, for simplicity because the $1$--body terms present in $2$--local Hamiltonians become less relevant to the random Hamiltonians that we consider as $n$ increases.  Their ellimination therefore allows us to estimate the scaling of the error and $\|H^{(n)}\|$ over a larger range of $n$.  

\begin{figure}
\centering
\includegraphics[width=0.8\textwidth]{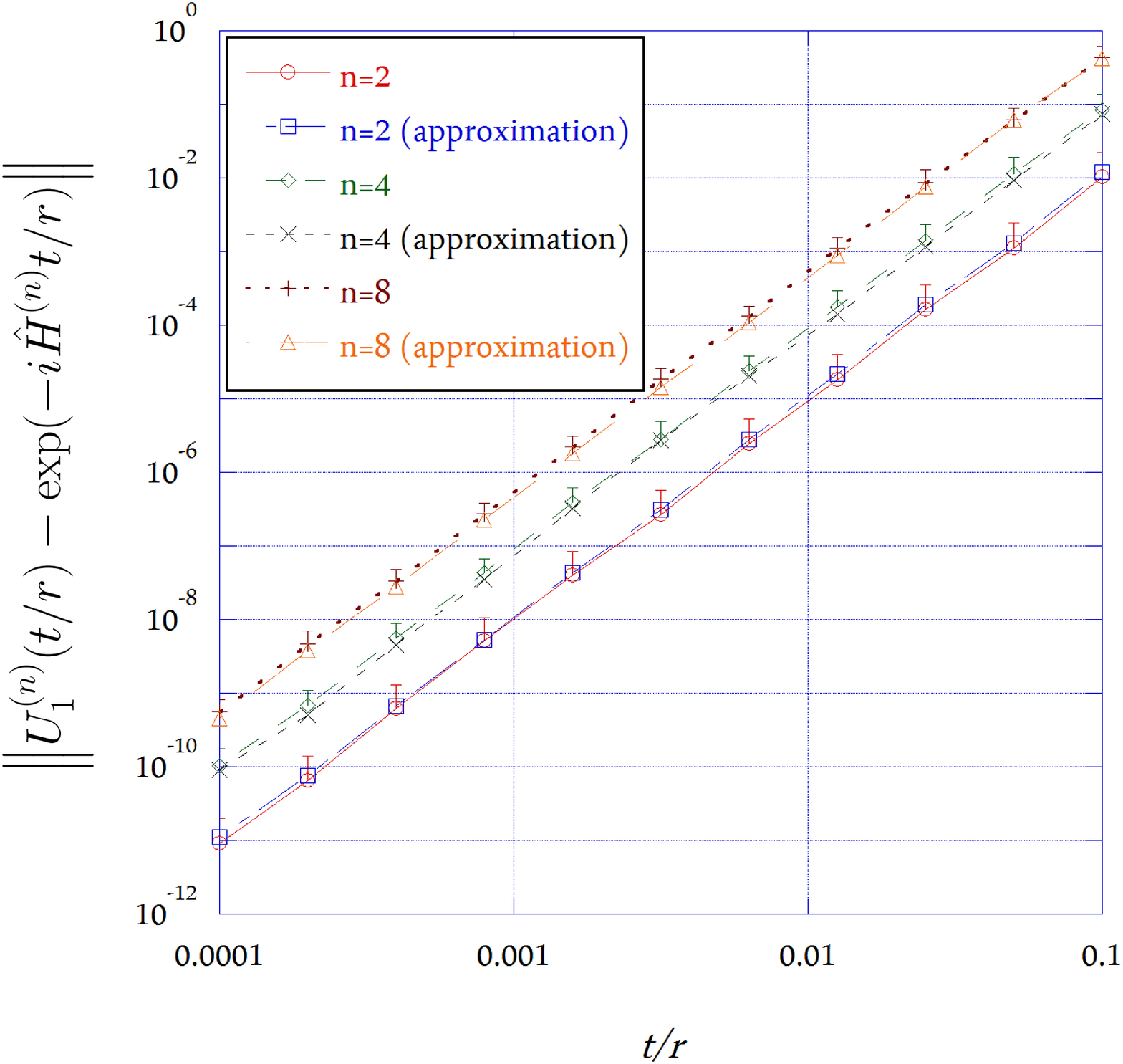}
\caption{
This plot shows the mean error in the lowest--order TS formula as a function of the duration of each time step and the approximation~(\ref{eq:ltserror})
where each point is randomly drawn from a Gaussian ensemble of $2$-body Hamiltonians.
Error bars are computed via the standard deviations of the measured results for the ensembles and only the upper bar is plotted because
the standard deviation is comparable or exceeds the mean value for all of these data points.  The data are therefore within statistical error of our
approximation.
\label{fig:TSerrork0}\protect}
\end{figure}

We estimate the error invoked by using the two lowest--order Trotter--Suzuki formulas
 by numerically sampling random $2$-body Hamiltonians taken from our Gaussian ensemble.    The
error in the formula is measured by the $2$-norm of the difference between the formula and the correct exponential for the randomly generated Hamiltonian.  We then repeat this for fifty different
randomly generated Hamiltonians for evolutions of duration $t\!=\!10^{-4}$ to $t\!=\!10^{-1}$ for $2$ through
$8$ qubits.  Higher--order integrators can be studied similarly, but low--order integrators are often the most significant for the current generation of experiments~\cite{BCC06}.
The error bound for the Strang--splitting (denoted $U_1^{(n)}$) in~\cite{WBHS10} gives
\begin{equation}
\left\|U^{(n)}_1\left(\frac{t}{r}\right)-\exp\left({-i\hat{H}^{(n)}\frac{t}{r}}\right)\right\|\le 2\left(\frac{3 m \max_{i} |a_i| t}{2r}\right)^3.\label{eq:ltsbd}
\end{equation}
We find from Fig.~\ref{fig:cmperror} that
the bound is, on average, too loose by approximately six orders of magnitude for $t\in [10^{-4},10^{-1}]$ for random two--body Hamiltonians acting on $n$ qubits.  This implies that much tighter estimates of the error in the Trotter--Suzuki formulas are needed to accurately estimate the performance of simulation algorithms.

Numerical estimates of the error are obtained by fitting the data in Fig.~\ref{fig:TSerrork0} that the average error for randomly sampled $2$--local Hamiltonians acting
on $2$--$8$ qubits is well modeled by
\begin{equation}
\left\|U^{(n)}\left(\frac{t}{r}\right)-U^{(n)}_1\left(\frac{t}{r}\right)\right\|\approx \frac{\left(\|\hat{H}^{(n)}\| \frac{t}{r}\right)^3}{3n^2},\label{eq:ltserror}
\end{equation}
for $U_1^{(n)}$ (the lowest order Trotter--Suzuki formula).  We see from the error bounds in Fig.~\ref{fig:TSerrork0} that the estimate of the error in $U_1(t/r)$ within an order of magnitude of the actual error for most of the randomly selected Hamiltonians.

Similarly, we
find by fitting polynomial functions to the data in Fig.~\ref{fig:TSerror} that 
\begin{equation}
\left\|U^{(n)}\left(\frac{t}{r}\right)-U_2^{(n)}\left(\frac{t}{r}\right)\right\|\approx \frac{1}{3000}\left(\frac{\|\hat{H}^{(n)}\| \frac{t}{r}}{\sqrt{n}}\right)^5,\label{eq:ltserror2}
\end{equation}
for random $2$--local Hamiltonians.
We can also find approximate forms for the error scaling for higher-order integrators, although doing so becomes
more difficult as numerical precision restricts the range of $t$ that can be used to assess the scaling.

\begin{figure}
\centering
\includegraphics[width=0.8\textwidth]{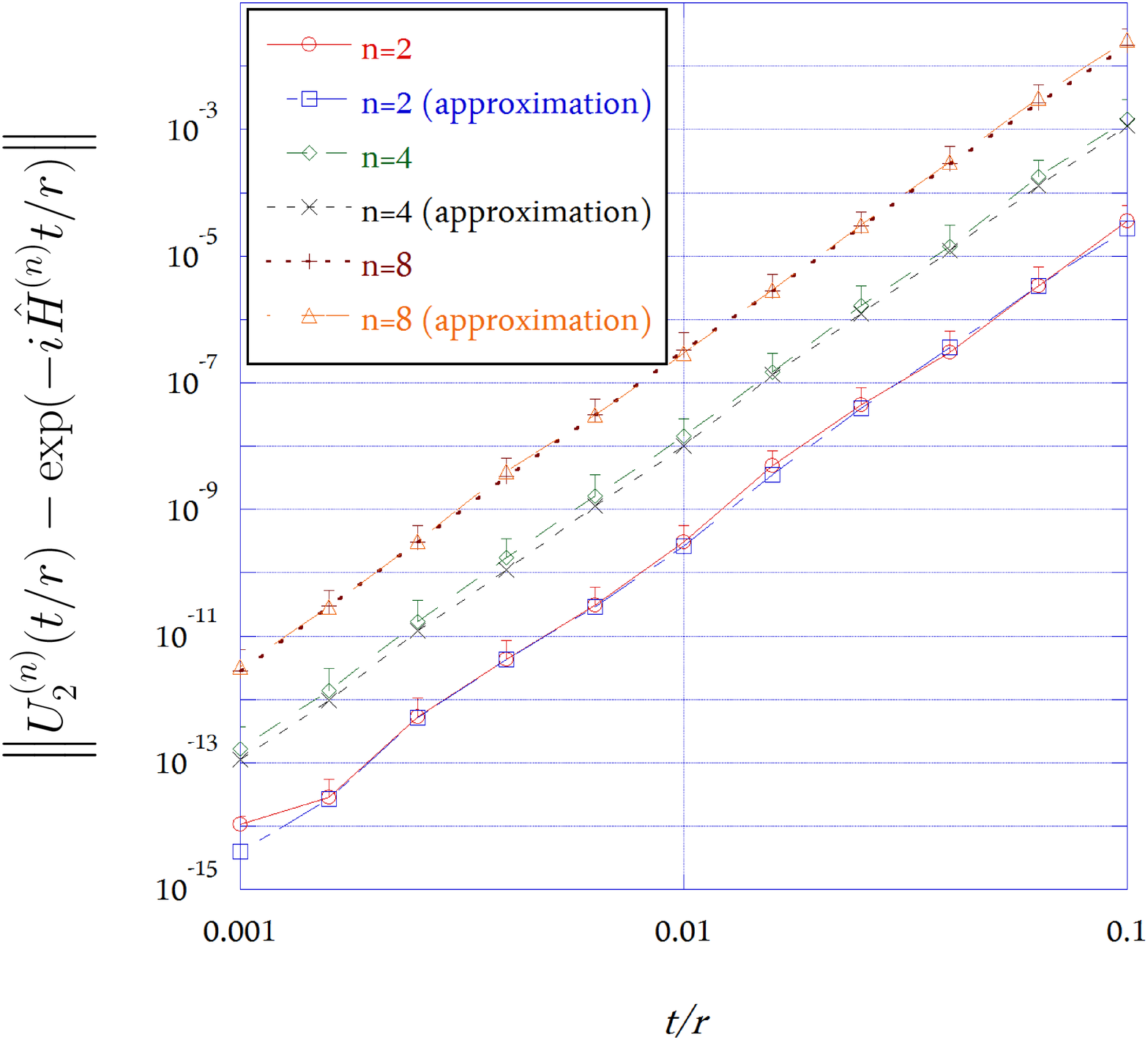}
\caption{
This plot shows the mean error in the second--lowest--order TS formula as a function of the duration of each time step and the approximation~(\ref{eq:ltserror2})
for a set of $50$ randomly generated $2$-local Hamiltonians sampled from our Gaussian ensemble for each data point.  We observe that the data is within statistical error of the fit in~(\ref{eq:ltserror2}).
\label{fig:TSerror}\protect}
\end{figure}

The total number of time steps needed to approximately simulate the Hamiltonian evolution within Trotter--error $\epsilon/2$
 using $U_1^{(n)}$ or $U_2^{(n)}$ follow directly from expressions~(\ref{eq:ltserror}) and~(\ref{eq:ltserror2}).
 The simulation errors are at most additive throughout the evolution, as each operator
in the product formula is unitary.  This implies that the total error is at most $\epsilon/2$ for
the case where $U_1^{(n)}$ is used if
\begin{equation}
r\left(\frac{(\|\hat{H}^{(n)}\| \frac{t}{r})^3}{3n^2} \right)\lessapprox \frac{\epsilon}{2}.
\end{equation}
Solving for $r$ gives the following approximate requirement for the number of time steps needed to satisfy the required
tolerance
\begin{equation}
r \gtrapprox \sqrt{\frac{2}{3n^2\epsilon}} \left({\|\hat{H}^{(n)}\|t} \right)^{1.5}.\label{eq:rscale1}
\end{equation}
The corresponding approximate requirement on $r$ for $U_2^{(n)}$ is
\begin{equation}
r \gtrapprox \frac{1}{30}\left(\frac{540}{n^{2.5}\epsilon}\right)^{1/4} \left({\|\hat{H}^{(n)}\|t} \right)^{1.25}.\label{eq:rscale2}
\end{equation}
These estimates of $r$ can be used in place of the default value in Algorithm~\ref{alg:main}; although we cannot rigorously guarantee that
they will provide error at most $\epsilon$.

Since there are $5$ times as many exponentials in $U_1^{(n)}(t/r)$ than there are exponentials in $U_2^{(n)}(t/r)$, inequalities~\eqref{eq:rscale1} and~\eqref{eq:rscale2} suggest that $U_1^{(n)}$ provides a shorter sequence of exponentials than the sequence consisting
of $r$ $U_2^{(n)}$ formul\ae~if 
\begin{equation}
\sqrt{\frac{2\left \|\hat{H}^{(n)}\right\|t}{3n^2\epsilon}} \le\frac{5}{30}\left(\frac{540\|\hat{H}^{(n)}\|t}{n^{2.5}\epsilon}\right)^{1/4},
\end{equation}
which is guaranteed if 
\begin{equation}\epsilon > \frac{16\left\|\hat{H}^{(n)}\right\| t }{15n^{3/2}}\label{eq:epsbd}.\end{equation}  
If inequality~(\ref{eq:epsbd}) is not satisfied, then higher--order formul\ae~such as $U_2^{(n)}$ are likely to yield more efficient simulation circuits.  Present simulation experiments, however, are often confined to short evolutions with relatively large error
tolerances (because direct comparison with numerical results is possible).  As a result, low--order approximations remain relevant for present experiments.

Tables~\ref{tab:1} and~\ref{tab:2} provide estimates of the number of exponentials that need to be implemented by Algorithm~\ref{alg:pauli}
in a simulation of a random $2$-body Hamiltonian.  We vary the number of qubits over an experimentally reasonable regime ($n=2,4,10$ qubits) and then extrapolate the scaling beyond the limitations of existing classical simulators to $40$ and $100$ qubits.  
In order to perform this extrapolation, we need to know the norm of the Hamiltonian.  We find from data fitting for cases up to $8$ qubits that the ensemble average of the norm of our random $2$--body Hamiltonians obeys
\begin{equation}
\|H^{(n)}\|\approx 1.3n^{5/3}.\label{eq:Hscale}
\end{equation}

Table~\ref{tab:1} shows how the number of required exponentials varies if $U_2$ is used instead of $U_1$ for a relatively modest value of $\epsilon$ and $t$, whereas Table~\ref{tab:2} presents the number of exponentials for a shorter evolution with a very small value of $\epsilon$.  Note that although the time is kept constant for these data sets, the (expected) norm of the Hamiltonian does not.  

\begin{table}[t]
\begin{minipage}[b]{0.4\linewidth}
\centering
\begin{tabular}{|c|c|c|c|}
\hline
n &$N_{\exp}$ for $U_1^{(n)}$&$N_{\exp}$ for $U_2^{(n)}$& Ratio\\
\hline
2 & 36	& 90 & 0.40\\
4 & 432 &540
&0.80\\
10 &8,190&8,100
&1.01 \\
40 &786,240&421,200
&1.87 \\
100 & 14,523,300&7,573,500
&1.92 \\
\hline
\end{tabular}
\caption{Extrapolated number of exponentials of $\hat{\mathfrak{h}}^{(n)}_j$ that have to be implemented to simulate a random $2$--body Hamiltonian with $a_j\sim \mathcal{N}(0,1)$ for $\epsilon=0.01$ and $t=0.1$.\label{tab:1}}
\end{minipage}
\hspace{1cm}
\begin{minipage}[b]{0.4\linewidth}
\centering
\begin{tabular}{|c|c|c|c|}
\hline
n &$N_{\exp}$ for $U_1^{(n)}$&$N_{\exp}$ for $U_2^{(n)}$&Ratio\\
\hline
2 & 108 &90
 &1.20\\
4 & 1,296&540
&2.40\\
10 &28,350&4,050
&7.00 \\
40 &2,471,040&280,800
&8.80 \\
100 &45,886,500&4,455,000
&10.3 \\
\hline
\end{tabular}
\caption{Extrapolated number of exponentials of $\hat{\mathfrak{h}}^{(n)}_j$ that have to be implemented to simulate a typical random $2$--body Hamiltonian with $a_j\sim \mathcal{N}(0,1)$ for $\epsilon=10^{-6}$ and $t=0.01$.\label{tab:2}}
\end{minipage}
\end{table}

The number of operations required in the simulation scales (for fixed order Trotter--Suzuki formulas) as $O(n^{2} r)$.  Our numerical estimate of the ensemble mean of the norm of $\|H\|$ in~\eqref{eq:Hscale} and the approximate bounds for $r$ in~\eqref{eq:rscale1} and~\eqref{eq:rscale2} lead us to the conclusion that the number of operations required to simulate a random $2$--body Hamiltonian scales with $n$ as $o(n^{3.4})$ as opposed to the $n^{4+o(1)}$ scaling predicted from the upper bounds for the error incurred by using either $U^{(n)}_1(t/r)$ or $U^{(n)}_2(t/r)$.  This suggests that the complexity of simulating random two--local Hamiltonians may be polynomially smaller than previous work implies.

The results in Tables~\ref{tab:1} indicates that using $U^{(n)}_1(t/r)$ instead of $U^{(n)}_2(t/r)$ leads to a reduction in the simulation complexity for small values of $n$, but $U_2^{(n)}(t/r)$ leads to more efficient simulations for $n\ge 40$.  In contrast, the data in Table~\ref{tab:2} shows that $U_2$ is more efficient than $U_1$ in every case considered except for the case where $n=2$.  We can therefore conclude that low--order Trotter--Suzuki formulas can sometimes be more efficient than high--order formulas for relatively undemanding simulation problems.  The data also suggests that extremely small gate errors (error on the order of $10^{-7}$ per gate) may be required to extend the DQS paradigm out to $40$ qubits and beyond because millions of gates are expected to be required for relatively modest simulations of $2$--local Hamiltonians.


\section{Conclusion}

In conclusion, we have designed an efficient classical algorithm for autonomous construction of efficient quantum circuits to simulate state evolution.
The circuits are costed in terms of a small standard gate set and require neither a Hamiltonian oracle nor ancillary qubits,
thereby making the universal quantum simulator minimal in space cost.
Furthermore, we show that the costing is still too pessimistic in some cases and could be improved but some orders of magnitudes. 
The algorithm also systematically searches out commuting terms in the Hamiltonian and groups them to
reduce the depths of the circuits yielded by our algorithm; thereby reducing the time required to execute the circuits using parallelism.

Our circuit construction algorithm is a significant advance because it is straightforward to implement the algorithm on a computer, it is highly efficient and it gives upper bounds for the simulation error.
Knowing error bounds is important for assessing the veracity of any quantum simulation.
Our work thus provides an important step towards constructing a practical, efficient and trustworthy simulator of quantum dynamics.

There are several remaining open problems that have not been addressed by this work.  First is the issue of simulating time--dependent quantum systems.
Such issues can be resolved by employing Trotter--Suzuki formul\ae~ for ordered operator exponentials~\cite{WBHS10,WBHS11} although the optimality of those error bounds for actual circuits remains to be checked using algorithmic methods we have introduced here.

Similarly, providing better upper bounds for $r$ remains an important problem as our work has
shown that in physically significant cases these bounds can be far too pessimistic.  Providing such bounds
would enable much more challenging simulation experiments to be performed in cases where rigorous error bounds are required of the output states.  

A final important extension of this work is discussing the optimization of the resulting circuits that are yielded by the algorithm.  Further optimization should be possible by using circuit identities to simplify the resulting circuits. Finding autonomous methods to optimize the output of such algorithms would be a significant asset in the development of a DQS that exceeds the power of existing classical simulators of quantum systems.

\appendix

\section{Condition for commutation of Pauli operators \label{appendix:pauli}}
Here we prove that~(\ref{eq:modeq}) gives a necessary and sufficient criterion for determining whether two Hamiltonians that
are tensor products of Pauli operators commute.   This criterion is essential to our discussion of parallelizing the quantum simulation because we need to divide the simulation into mutually commuting 
sections in order to exploit parallelism.

We simplify our discussion by removing from $\hat{\mathfrak{h}}_i^{(n)}$ and $\hat{\mathfrak{h}}_j^{(n)}$ every qubit that is either acted on as the identity operator by at least one of the Hamiltonians or every qubit that both Hamiltonians have the same action upon.  These qubits are removed from consideration because they are not needed to determine the commutation properties.  
After removing these irrelevant qubits, we simplify the discussion by relabeling the remaining qubits to group them in six different groups.  These simplifications lead to the following representation for $\hat{\mathfrak{h}}_i^{(n)}$
\begin{eqnarray}
\hat{\mathfrak{h}}_{i}^{(n)}&\sim& X^{\otimes |\bS_{xi}\bigcap\bm{S}_{yj}|}\otimes Y^{\otimes |\bS_{yi}\bigcap\bm{S}_{zj}|}\otimes Z^{\otimes |\bS_{zi}\bigcap\bm{S}_{xj}|}\nonumber\\ &~&\otimes X^{\otimes |\bS_{xi}\bigcap\bm{S}_{zj}|}\otimes Y^{\otimes |\bS_{yi}\bigcap\bm{S}_{xj}|}\otimes Z^{\otimes |\bS_{zi}\bigcap\bm{S}_{yj}|}.
\end{eqnarray}
The simplified expression for $\hat{\mathfrak{h}}_j^{(n)}$ is
\begin{eqnarray}
\hat{\mathfrak{h}}_{j}^{(n)}&\sim& Y^{\otimes |\bS_{xi}\bigcap\bm{S}_{yj}|}\otimes Z^{\otimes |\bS_{yi}\bigcap\bm{S}_{zj}|}\otimes X^{\otimes |\bS_{zi}\bigcap\bm{S}_{xj}|}\nonumber\\ &~&\otimes Z^{\otimes |\bS_{xi}\bigcap\bm{S}_{zj}|}\otimes X^{\otimes |\bS_{yi}\bigcap\bm{S}_{xj}|}\otimes Y^{\otimes |\bS_{zi}\bigcap\bm{S}_{yj}|}.
\end{eqnarray}

We can then evaluate the product of the two operators using the property that $XY=iZ$, $YZ=iX$ and $ZX=iY$ and also using
the anti--commutativity of Pauli--operators.  This implies
\begin{eqnarray}
\hat{\mathfrak{h}}_{i}^{(n)}\hat{\mathfrak{h}}_{j}^{(n)}&\sim& (iZ)^{\otimes |\bS_{xi}\bigcap\bm{S}_{yj}|}\otimes (iX)^{\otimes |\bS_{yi}\bigcap\bm{S}_{zj}|}\otimes (iY)^{\otimes |\bS_{zi}\bigcap\bm{S}_{xj}|}\nonumber\\ &~&\otimes (-iY)^{\otimes |\bS_{xi}\bigcap\bm{S}_{zj}|}\otimes (-iZ)^{\otimes |\bS_{yi}\bigcap\bm{S}_{xj}|}\otimes (-iX)^{\otimes |\bS_{zi}\bigcap\bm{S}_{yj}|}.\label{eq:appendixb:ij}
\end{eqnarray}
Conversely,
\begin{eqnarray}
\hat{\mathfrak{h}}_{j}^{(n)}\hat{\mathfrak{h}}_{i}^{(n)}&\sim& (-iZ)^{\otimes |\bS_{xi}\bigcap\bm{S}_{yj}|}\otimes (-iX)^{\otimes |\bS_{yi}\bigcap\bm{S}_{zj}|}\otimes (-iY)^{\otimes |\bS_{zi}\bigcap\bm{S}_{xj}|}\nonumber\\ &~&\otimes (iY)^{\otimes |\bS_{xi}\bigcap\bm{S}_{zj}|}\otimes (iZ)^{\otimes |\bS_{yi}\bigcap\bm{S}_{xj}|}\otimes (iX)^{\otimes |\bS_{zi}\bigcap\bm{S}_{yj}|}.\label{eq:appendixb:ji}
\end{eqnarray}

We then see from~(\ref{eq:appendixb:ij}) and~(\ref{eq:appendixb:ji})  that $[\hat{\mathfrak{h}}_{i}^{(n)},\hat{\mathfrak{h}}_{j}^{(n)}]=0$ if and only if
\begin{eqnarray}
 |{\bm S_y}_i \cap {\bm S_x}_j|\!+\!|{\bm S_x}_i \cap {\bm S_z}_j|\!+\!|{\bm S_z}_i \cap {\bm S_x}_y|\!\equiv\!  
|{\bm S_x}_i \cap {\bm S_y}_j|\!+\!|{\bm S_z}_i \cap {\bm S_x}_j|\!+\!|{\bm S_y}_i \cap {\bm S_z}_y|~({\rm mod}~ 2),\nonumber\\\label{eq:appendixb:modeq2}
\end{eqnarray}
which is equivalent to~(\ref{eq:modeq}).

\acknowledgements
We acknowledge MITACS, USARO, NSERC, and AITF for financial support and thank M. M\" uller for helpful comments.
BCS is supported by a CIFAR Fellowship.

\bibliography{qsim}

\end{document}